# MOSAIC on the ELT: High-multiplex Spectroscopy to Unravel the Physics of Stars and Galaxies from the Dark Ages to the Present Day

François Hammer[1]
Simon Morris[2]
Jean-Gabriel Cuby[3]
Lex Kaper[4,18]
Matthias Steinmetz[5]
Jose Afonso[6]
Beatriz Barbuy[7]
Edwin Bergin[8]
Alexis Finogenov[9]
Jesus Gallego[10]
Susan Kassin[11]
Chris Miller[8]
Goran Östlin[12]
Laura Pentericci[13]
Daniel Schaerer[14]
Bodo Ziegler[15]
Fanny Chemla[1]
Gavin Dalton[16]
Fatima De Frondat[1]
Chris Evans[17]
David Le Mignant[3]
Mathieu Puech[1]
Myriam Rodrigues[1]
Ruben Sanchez-Janssen[17]
Sylvestre Taburet[1]
Lidia Tasca[3]
Yanbin Yang[1]
Sandrine Zanchetta[1]
Kjetil Dohlen[3]
Marc Dubbeldam[2]
Kacem El Hadi[3]
Annemieke Janssen[18]
Andreas Kelz[5]
Marie Larrieu[19]
Ian Lewis[20]
Mike MacIntosh[17]
Tim Morris[2]
Ramon Navarro[18]
Walter Seifert[21]

[1] Paris Observatory, Paris Science et Lettres, CNRS, France
[2] Durham University, UK
[3] LAM, Université Aix-Marseille, CNRS, France
[4] University of Amsterdam, the Netherlands
[5] Leibniz-Institut für Astrophysik Potsdam, Germany
[6] IACE, Universidade de Lisboa, Portugal
[7] IAG, São Paulo, Brazil
[8] University of Michigan, USA
[9] University of Helsinki, Finland
[10] Universidad Complutense de Madrid, Spain
[11] Space Telescope Science Institute, Baltimore, USA
[12] Stockholm University, Sweden
[13] INAF – Osservatorio Astronomico di Roma, Italy
[14] University of Geneva, Switzerland
[15] Vienna University, Austria
[16] STFC-RALSPACE & University of Oxford, UK
[17] UK Astronomy Technology Centre, STFC, Edinburgh, UK
[18] NOVA, the Netherlands
[19] IRAP, Université de Toulouse, CNRS, France
[20] University of Oxford, UK
[21] LSW Heidelberg, Germany

The powerful combination of the cutting-edge multi-object spectrograph named MOSAIC with the world largest visible/near-infrared telescope, ESO's Extremely Large Telescope (ELT), will allow us to probe deeper into the Universe than ever before. MOSAIC is an extremely efficient instrument for obtaining spectra of the numerous faint sources in the Universe, including the very first galaxies and sources of cosmic reionisation. MOSAIC has a high multiplex in the near-infrared (NIR) and in the visible, and also has multi-integral field units (Multi-IFUs) in the NIR. It is therefore perfectly suited to carrying out an inventory of dark matter (from rotation curves) and baryons in the cool–warm gas phases in galactic haloes at $z = 3–4$. MOSAIC will enable detailed maps of the intergalactic medium at $z = 3$, the evolutionary history of dwarf galaxies during a Hubble time, and the chemistry as directly measured from stars up to several Mpc. It will also measure faint features in cluster gravitational lenses or in streams surrounding nearby galactic haloes. The preliminary design of MOSAIC is expected to begin next year and its level of readiness is already high, given the instrumental studies already carried out by the team.

## Science cases for MOSAIC

The science case for a multi-object spectrograph on the ELT covers all areas of astronomy for which the collection of light from tens to hundreds of scientific targets is required. This includes targets which are too faint to be accessible spectroscopically to 10-m telescopes, particularly stars, star clusters, and all types of galaxies across cosmic history. As a major workhorse for the ELT, MOSAIC will provide access to the visible and NIR wavelength ranges.

Access to visible wavelengths is an important priority for the large community of scientists interested in collecting numerous stellar spectra in the Milky Way and nearby galaxies (see, for example, Evans et al., 2013, 2015). In cosmology, this enables maps of the intergalactic medium (IGM) and also the determination of galactic halo abundances at $z = 3–4$ from absorption lines in the spectra of the numerous Lyman break galaxies (Japelj et al., 2019; Rahmani et al., in preparation). In the NIR, MOSAIC will constrain the physics of early star formation by determining the metallicity distribution functions of stellar populations in the local Universe, and it will excel at deep spectroscopic studies of faint Galactic satellites — from the chemodynamics of ultra-faint dwarf galaxies to detailed chemical abundances of old globular clusters in the inner Galaxy via high-resolution spectroscopy. Furthermore, MOSAIC's unique system of multi-IFUs will enable measurements of the rotation curves of distant galaxies. With such observations MOSAIC will lead the efforts to characterise and understand baryonic and dark matter structure at an epoch when the Universe was only 1.5 billion years old. Looking into the even more distant Universe, MOSAIC's multi-IFUs will study the physical properties of the stars, interstellar medium (ISM), radiation field, and ionising power of the overall population of faint very high-$z$ galaxies. This will enable a precise characterisation of the ionisation state of the IGM during the first billion years after the Big Bang. MOSAIC will allow astronomers to reconstruct the timeline and topology of reionisation, and to observe the formation and growth of the first galaxies in much more detail than any other instrument.

The MOSAIC Science Team has collected science cases that should revolutionise astrophysics and cosmology in the 2030s. It has also defined a number of large surveys that could be carried out with MOSAIC (Evans et al., 2018; Puech et al., 2018) to illustrate the survey speed of the instrument. Although subject to revision during the construction phase, these exciting science cases span a range of redshift from our "backyard" to the dark ages:
– First-light galaxies and reionisation (with multi-IFUs in the NIR, and the high-multiplex mode in the NIR for Lyman-α emitters).
– An inventory of the matter distribution at large scales, including tomography of the IGM and its metal content (with the high-multiplex mode in the visible) and dark matter distributions from rotation curves of distant galaxies (with multi-IFUs in the NIR).
– Mass assembly and evolution of dwarf galaxies (with the high-multiplex mode in the visible and NIR).
– Extragalactic stellar populations (with the high-multiplex mode in the visible and multi-IFUs in the NIR).

## Instrument concept

The MOSAIC instrument was conceived as a multi-purpose multi-object spectrograph for the ELT (Hammer et al., 2016; Morris et al., 2018). It has three main operating modes, including two high-multiplex mono-aperture modes (in the visible and NIR) and the multi-IFU mode, as shown in Figure 1. The requirement to achieve a multiplex of ~ 200 in the high-multiplex mode with a multi-IFU mode essentially predetermines that MOSAIC shall be a fibre-fed instrument. Furthermore, the design is considerably simplified by allowing the spectrograph hardware to be shared between the different modes. The spectral range of the instrument is from 0.45 to 1.8 μm, with a break between a system optimised for visible and a system optimised for the NIR at 0.8–0.9 μm. Six bands provide $R = 5000$ coverage over the full bandwidth, with selected smaller bands available at $R \sim 20\,000$ in the visible and in the NIR. The instrument is designed to use as much of the adaptive-optics-corrected field of view of the ELT as possible (> 7 arcminutes in diameter). It is the only currently planned instrument that takes advantage of the wide ELT field of view with high optical quality enabled by its 5-mirror telescope design.

The design shown in Figure 2 has been adopted to achieve an acceptable resistance to gravity-induced deformations on the platform. MOSAIC will include two spectrographs in the visible and four in the NIR. Owing to the large plate scale offered by the ELT (> 3 mm arcsecond$^{-1}$) and the non-telecentricity of the telescope optical design, a number of compromises are required in order to realise a practical and affordable multi-object spectrograph instrument design. In particular, it has been necessary to subdivide individual pick-offs into multiple fibre-spaxels in order to produce an optical design that is feasible. A wider simultaneous spectral coverage is appealing, but it has to be balanced against the need to keep the spaxel size sufficiently small. This is to ensure the leading role of MOSAIC in measuring the complex kinematics of high-$z$ galaxies (for example, Hammer et al., 2009; Kassin et al., 2012) and detecting spectral features in the faintest sources, including the first galaxies.

## Performance

### Studies in the visible wavelength range

The ELT's light-gathering power will exceed that of all sixteen currently existing 8–10-m-class telescopes put together. In the visible, MOSAIC will provide world-leading science by producing ≥ 200 spectra simultaneously for Galactic and nearby Universe science. Studies of the Local Group have shown that precise chemical abundances and stellar kinematics can break the age-metallicity degeneracy. MOSAIC will bring a wealth of new and exciting target galaxies within our grasp for the first time, spanning a much broader range of galaxy morphologies, star formation histories and metallicities than those available to us at present and enabling us to test theoretical models of galaxy assembly and evolution in systems out to distances of several or even tens of Mpc. Another key objective in this area is also to investigate the existence of the "Spite Plateau" (Spite & Spite, 1982) in the Li abundances of metal-poor stars in extragalactic systems. Because the observed Li is likely primordial, the Spite Plateau provides an estimate of the baryonic density of the Universe. This case requires high-resolution spectroscopy ($R > 20\,000$).

For cosmology, the visible mode will enable a detailed mapping of the IGM at $z = 3$–4. Figure 3 shows simulations indicating that MOSAIC will be able to measure Lyman-α forest tomography, even with relatively faint background Lyman break galaxies at z ~ 4, as shown by Japelj et al. (2019). By measuring the optical depths of redshifted ultraviolet metallic absorption lines, one may also characterise the metal content in the haloes of z = 3–4 galaxies (Rahmani et al., in preparation). That will yield a comprehensive inventory of baryons in the distant haloes, which can be compared to what is presently done in nearby haloes by the HST-COS Halo collaboration (see, for example, Werk et al., 2014).

Figure 1. The three observing modes of MOSAIC. Left: the two high-multiplex modes (HMMs); right: the 10 multi-IFUs.

| High multiplex mono-fibres | Visible | Near IR | Multiple integral field units | Near IR |
|---|---|---|---|---|
| 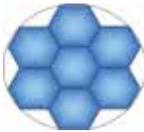 | 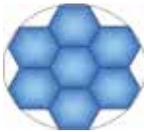 | 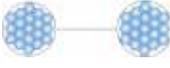 | | 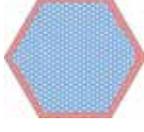 |
| Number of apertures | 220 | 160 (80 sci + 80 sky) | Number of apertures | 10 |
| Patrol area | 52.1 arcmin$^2$ | 47.3 arcmin$^2$ | Patrol area | 44.2 arcmin$^2$ |
| Operating bandwidth | 0.45–0.88 μm | 0.8–1.8 μm | Operating bandwidth | 0.8–1.8 μm |
| Diameter of the aperture on sky | 690 mas | 500 mas | Outer diameter of on-sky subfield | 2.5 arcsec (hexagonal) |
| Spectral resolution | 5000 & 20 000 | 5000 & 20 000 | Sampling | 120 mas |
| AO performance | GLAO (~seeing limited) | GLAO | Spectral resolution | 5000 & 20 000 |
| | | | AO performance | 25% encircled energy in 150 mas |

First galaxies and cosmic reionisation in the NIR wavelength range

The ELT's light-gathering power will allow us to study the physical properties of the faintest sources ever reached by medium- or high-resolution spectroscopy, including the very first sources in the Universe, which are barely detected today. These objects are challenging because they fall into the category of ultra-low surface brightness sources (ULSBs, $\mu_{AB} \geq 25$ mag arcsecond$^{-2}$) as a result of cosmological dimming. Whereas the James Webb Space Telescope (JWST) will excel at very deep NIR and mid-IR imaging, and will measure for the first time the strong rest-frame optical emission lines for large numbers of high-z galaxies, it will not be able to perform moderate-resolution ($R \geq 5000$) spectroscopy, which is essential for measuring the absorption and faint emission lines of these distant sources. Combining the ELT's large collecting area with multi-IFUs to optimise sky subtraction (cf. below), MOSAIC will allow NIR spectroscopy significantly beyond the reach of the JWST (cf. Vanzella et al., 2014).

As regards studying the sources of cosmic reionisation, the JWST is in practice limited by the fact that so far no method has been found to measure the Lyman continuum escape fraction of galaxies from their rest-frame optical emission lines (for example, Plat et al., 2019; Wang et al., 2019; Ramambason et al., 2020), whereas the rest-frame ultraviolet spectra admit several methods to determine this fundamental quantity (Izotov et al., 2018; Chisholm et al., 2018, 2020). Observing the ultraviolet rest-frame continuum and faint emission lines of a large sample of z > 6 galaxies, which is uniquely feasible with the ELT, is crucial for understanding cosmic reionisation and the first sources of light, for several reasons:

– Low-ionisation absorption lines (for example, Si II, C II) provide a unique tool from which to derive the Lyman continuum escape fraction and to determine the neutral gas coverage in these sources (Steidel et al., 2018; Chisholm et al., 2018; Gazagnes et al., 2018). This is essential in order to establish the contribution of galaxies to cosmic reionisation, and to determine which galaxy types dominate.
– Measurements of Lyman-α emission (which is generally fairly weak at z > 6.5), the detailed line profiles, and statistics (for example, the fraction of galaxies showing Lyman-α emission) provide critical information about the ionisation state of the IGM, the timeline of cosmic reionisation, and its topology (for example, Castellano et al., 2016; Mason et al., 2018; Mesinger et al., 2015).
– Other emission lines (such as C III], O III], C IV, He II etc.) provide important insight into the ISM, stellar metallicity and massive star populations, and the radiation field of these galaxies. In particular, MOSAIC will enable the search for primordial, metal-free star-formation for an unprecedented sample of galaxies through the He II emission line (for example, Raiter, Schaerer & Fosbury, 2010). The rest-frame ultraviolet emission lines are weak and therefore require ultra-deep spectroscopy (Stark, 2016).

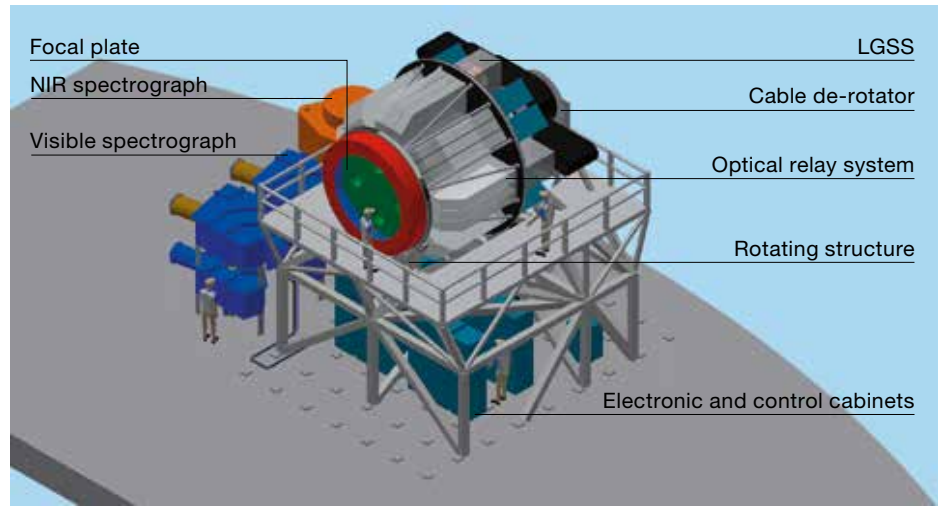

Figure 2. Schematic of MOSAIC as it will be on the ELT Nasmyth platform.

High multiplex with mono-aperture fibres in the NIR will be very useful for investigating bright Lyman-α line sources at high redshift. However, for the numerous very distant and intrinsically faint galaxies that have no strong emission, MOSAIC will be unique in its ability to measure continuum, absorption lines and very faint emission lines. The effective depth provided by the ELT's light-gathering power is affected by the strong and variable sky signal, and even for spectroscopic studies operating between the bright OH sky lines, the NIR sky continuum $\mu_{AB}$ = 18.5–19.5 mag arcsecond$^{-2}$ (see Sullivan & Simcoe, 2012) is far brighter than the typical surface brightness associated with the continuum of very distant galaxies at $\mu_{AB} \geq 25$ mag arcsecond$^{-2}$. Achieving the best sky subtraction is therefore a major driver for the design of a multi-object spectrograph on the ELT. Weilbacher et al. (2020) have recently presented detailed results from the Multi Unit Spectroscopic Explorer (MUSE) IFU, demonstrating that it is capable of reaching a dark sky removal down to 0.3% between 0.82 and 0.9 μm. Figure 4 compares the mono-aperture fibre performance to that of IFUs (see also Disseau et al., 2014). To recover signals well below the sky, we estimate that an IFU could be 2 to 3 times more efficient than mono-aperture fibres, after taking account of the sky spatial variations near and far from the target (see, for example, Yang et al., 2012), the sky temporal variations, and the expected aperture losses. Multi-IFUs are essential for recovering spectra of most $z \sim 8$ galaxies, as illustrated in the right-hand panel of Figure 4.

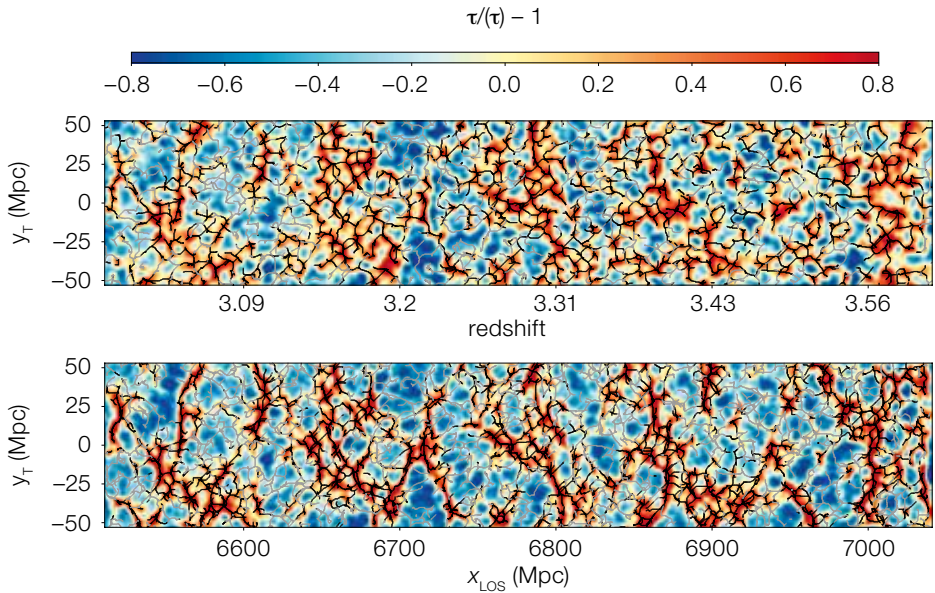

Figure 3. Simulations of MOSAIC performance when observing simultaneously the ISM absorption lines in the spectra of numerous $R_{AB} \leq 25.5$ Lyman break galaxies enabling a vast area (100 Mpc in diameter) of the sky plane to be sampled. This allows the study of 10-Mpc-thick longitudinal slices lying from redshift 3 to 3.6, of the simulated (top, signal-to-noise ratio = 4) and original (bottom) optical depth contrast field ($\tau/\langle\tau\rangle - 1$), where $\tau = -\log F$ and F is transmitted flux ($\tau$ is taken as a proxy for the HI density). The y-axis is the projected transverse dimension ($y_T$). Filaments extracted are over-plotted in grey and black. The black lines correspond to the 50% densest filaments. As noise in the spectra increases, the reconstruction shows lower contrast, and filaments are more randomly located (from Japelj et al., 2019).

## Comparison with other instruments and telescopes

Puech et al. (2018) carried out a systematic comparison between the efficiency of MOSAIC and that of other facilities implemented on the ELT and 8–10-m-class telescopes. Expressed in a survey speed metric — for example, how fast a given instrument can observe a given number of sources to equivalent signal-to-noise ratio levels — it demonstrates that only MOSAIC on the ELT will be able to construct large statistical samples in reasonable time and to address the science cases described in the first section of this article. Interestingly, the JWST will serve to identify many ultra-faint targets that MOSAIC will follow up to derive their physical, chemical, and kinematical properties. MOSAIC also nicely complements the High Angular Resolution Monolithic Optical and Near-infrared Integral field spectrograph (HARMONI), whose largest spaxel scale is 3 times smaller than that of MOSAIC. HARMONI will excel at studying compact sources or targets with high-contrast small-scale clumps while MOSAIC will be optimised to study the large number of $z > 6$ galaxies, which have relatively small sizes but are extended (Bowler et al., 2017) and are in the range of ultra-low-surface-brightness sources.

## Conclusions

In addition to the four scientific topics listed in the first section of this article, as a workhorse instrument MOSAIC will be able to address a huge number of science cases. The Science Team is presently developing many new science cases, including, for example, kinematics and stellar populations of faint stellar streams surrounding nearby galaxies (Martínez-Delgado, 2018), velocity dispersion measurements of dwarf spheroidals surrounding galactic haloes situated up to 50 Mpc from us (as done by MUSE on Crater I, see Voggel et al., 2016), and many others. The large MOSAIC science community will continue to prepare innovative and cutting-edge observational programmes for studying stars and galaxies in many environments and across cosmic history.

The construction phase of MOSAIC could begin in 2021. This has been folded into the planning by the MOSAIC team to ensure we deliver an instrument that will have no equal when it comes to exploring the deep Universe towards the epoch of reionisation. MOSAIC will be a unique instrument with which to investigate how the gas, stars, and dark matter were distributed more than 10 billion years ago, to determine the properties of the faintest building blocks of galaxies, and to provide a gigantic step towards understanding the local Universe well beyond the Local Group. In short, it is the ideal instrument with which to investigate the physical properties of statistical samples of the vast majority of extragalactic sources because they are extended and have very low surface brightness. The MOSAIC team remains very keen to work with the ELT community to deliver a world-class instrument serving as many scientific interests as possible.

### Acknowledgements

The MOSAIC Consortium is very grateful to the MOSAIC Science Team[1] for their considerable help in envisioning the future multi-object spectrograph for the ELT. We are indebted to the whole MOSAIC technical team whose efforts are invaluable for preparing and conceiving MOSAIC and making it a reality. Special thanks go to Thierry Botti (LAM) and Frederique Auffret (Paris Observatory) for their help within the communication team, and to Clotilde Laigle and Jure Japelj for sending us Figure 3 in original form.

This article is dedicated to the memory of Olivier Le Fèvre, who contributed so brilliantly to the science discussed here. The MOSAIC Consortium is very grateful for his crucial support over the years.

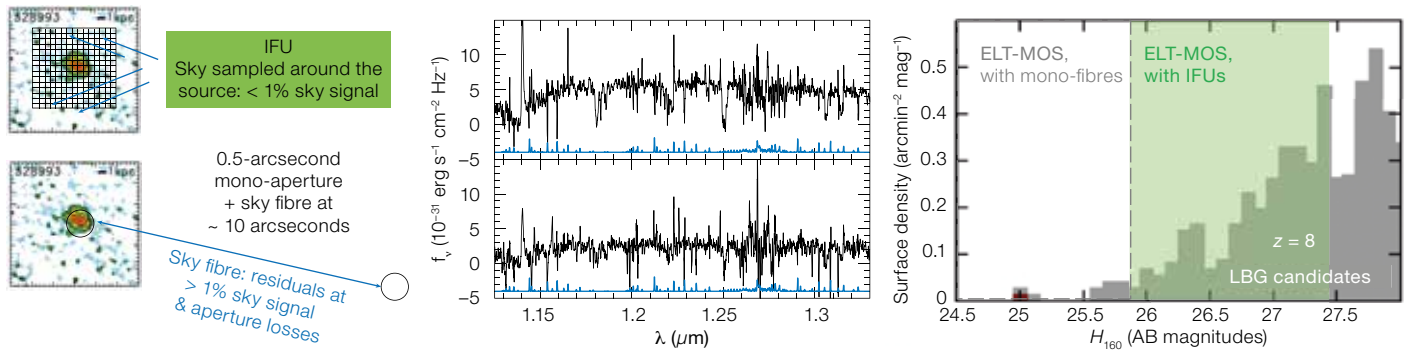

Links

[1] Science Team list: http://www.mosaic-elt.eu/index.php/science/46-science-team

Figure 4. Comparison between IFU and mono-aperture performance, illustrating the necessity of MOSAIC's multiple observing modes. Left: Sky subtraction schemes for the two modes. Middle: Spectra of SGASJ122651.3+215220 (see Rigby et al., 2018) redshifted to $z = 8.38$ and $AB = 27$. The top spectrum has been extracted using WEBSIM (Puech et al., 2016) assuming an IFU on the ELT (30 hrs exposure), providing a signal-to-noise ratio of $\sim 30$ for the absorption lines. The bottom spectrum extraction is the same but made with a 0.5-arcsecond fibre aperture and a sky correction uncertainty of 1%, showing that it does not allow absorption line measurements. A similar result would be obtained by using 0.5% and 1.2% sky residuals for an IFU and mono-aperture fibre, respectively. Right: Adapted from Figure 4 of Oesch et al. (2015), this shows the surface density of the full sample of $z \sim 8$ galaxies in the deep fields studied by Bouwens et al. (2015) (grey histogram). The green area shows the magnitude range of Lyman break galaxy (LBG) candidates that can be targeted with the multi-IFUs of MOSAIC, thanks to their better throughput when observing low-surface-brightness sources.

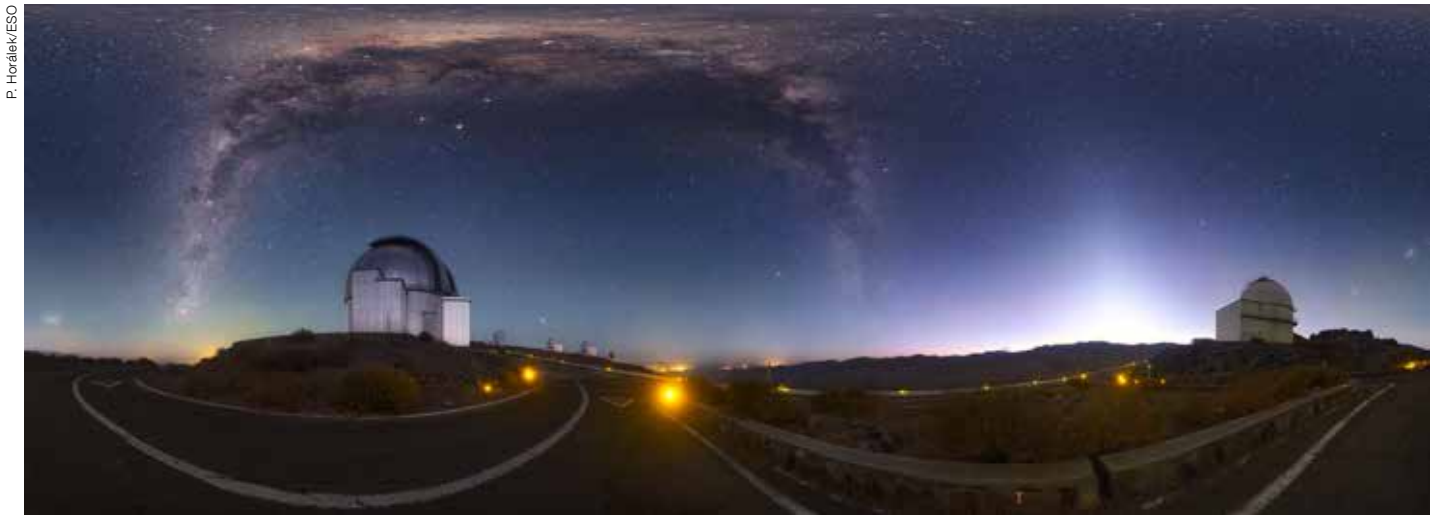

The arrival of daylight at ESO's La Silla Observatory reveals the splendour of the Universe beyond our little planet in dazzling detail. The Milky Way stretches overhead as a streaming banner of dust backlit by the light of billions of stars. Clouds of interstellar dust grow thickest towards the constellation of Sagittarius (The Archer), which marks the centre of the galaxy — the core around which the spectacular spiral arms rotate.